\documentclass{aa}  

\usepackage{graphicx}
\usepackage{multirow}
\usepackage{soul}
\usepackage{cancel}

\usepackage{txfonts}
\usepackage[dvipsnames]{xcolor}
\usepackage{orcidlink}

\begin{document}
\renewcommand{\sectionautorefname}{Section}
\renewcommand{\subsectionautorefname}{Subsection}

   \title{Revisiting the angular size--redshift cosmological test with milliarcsecond radio structures in active galactic nuclei}

\titlerunning{Revisiting the angular size--redshift cosmological test}

 \author{Mina Ghodsi Yengejeh\inst{1,2,3,4}\orcidlink{0000-0001-5481-9810}
            \and
        Tatiana A. Koryukova\inst{5}\orcidlink{0000-0001-8347-7880}
            \and
        Leonid I. Gurvits\inst{6,7}\orcidlink{0000-0002-0694-2459} 
          \and
        S\'andor Frey\inst{2,3,4}\orcidlink{0000-0003-3079-1889}
          \and \\
        Alexander B. Pushkarev\inst{8,9}\orcidlink{0000-0002-9702-2307}
          \and
        Alexander V. Plavin\inst{10}\orcidlink{0000-0003-2914-8554}
          \and
        Kenneth I. Kellermann\inst{11}\orcidlink{0000-0002-0093-4917}
          \and
        Andr\'as Kov\'acs\inst{1,2,3}\orcidlink{0000-0002-5825-579X}
          }
\authorrunning{M. Ghodsi Yengejeh et al.}

   \institute{
            MTA--CSFK  \emph{Lend\"ulet} ``Momentum'' Large-Scale Structure (LSS) Research Group, Konkoly Thege Mikl\'os \'ut 15-17, H-1121 Budapest, Hungary
        \and
            Konkoly Observatory, HUN-REN Research Centre for Astronomy and Earth Sciences, Konkoly Thege Mikl\'os \'ut 15-17, H-1121 Budapest, Hungary
         \and
            CSFK, MTA Centre of Excellence, Konkoly Thege Mikl\'os \'ut 15-17, H-1121 Budapest, Hungary
        \and
            Institute of Physics and Astronomy, ELTE E\"otv\"os Lor\'and University, P\'azm\'any P\'eter s\'et\'any 1/A, H-1117 Budapest, Hungary
        \and
            Lebedev Physical Institute of the Russian Academy of Sciences, Leninsky prospekt 53, 119991 Moscow, Russia
        \and
            Joint Institute for VLBI ERIC, Oude Hoogeveensedijk 4, 7991 PD Dwingeloo, The Netherlands
         \and
            Faculty of Aerospace Engineering, Delft University of Technology, Kluyverweg 1, 2629 HS Delft, The Netherlands
        \and
            Crimean Astrophysical Observatory, Nauchny 298409, Crimea
        \and
            Lebedev Physical Institute, Astro Space Center, Pushchino Radio Astronomy Observatory, Radiotelescopnaya 1a, Pushchino 142290, Russia
        \and
            Black Hole Initiative at Harvard University, 20 Garden Street, Cambridge, MA 02138, USA
        \and
            National Radio Astronomy Observatory, 520 Edgemont Rd., Charlottesville, VA 22903, USA
                }

   \date{Received 30 August 2025; Accepted 13 April 2026}

  \abstract
   {Very long baseline interferometry (VLBI) measurements of the sizes of compact extragalactic radio sources, jetted active galactic nuclei, provide data for probing the angular size--redshift relation, offering a classical cosmological test complementary to other distance--redshift methods.}
   {Aiming to update and extend previous studies conducted in the 1990s, we analyse a significantly expanded and improved dataset to reassess the angular size--redshift relation and its potential for constraining cosmological model parameters, focusing on the matter density parameter $\Omega_{\mathrm{m}}$ in a flat $\Lambda$ Cold Dark Matter Universe. This is the first major update of the compact-source angular size test in the past quarter of a century, using a dataset an order of magnitude larger than in previous studies.}
   {We performed a Markov chain Monte Carlo (MCMC) analysis on real data and on multiple mock catalogues with varying Gaussian noise levels ($10\%,\,20\%,\, 50\%$) to evaluate parameter constraints in the presence of observational scatter. In addition, we conducted a test with 100 randomized catalogues created by shuffling redshifts while preserving other observables to explore the statistical significance of the angular size--redshift dependence. We also explored how astrophysical parameters depend on fixed cosmological models with different $\Omega_{\mathrm{m}}$ values.}
   {The randomization test showed that the posterior distributions from randomized data do not overlap with those from real observations, with significant deviations, confirming that the measured angular size--redshift relation is physically meaningful and not a chance alignment. The astrophysical model parameter that describes the redshift dependence of the source angular size exhibits strong sensitivity and degeneracy with $\Omega_{\mathrm{m}}$. Simulated mock catalogues indicate that the method is able to constrain $\Omega_{\mathrm{m}}$ if the data scatter is below $\sim20\%$, but current real data noise levels are too high for reaching competitive cosmological constraints. Scaling estimates suggest that high-quality data of samples of several thousands to $\sim 100\,000$ sources, a standardisation calibration approach, and/or refining sample selection criteria are needed to fully exploit the potential of the angular size--redshift test with this type of objects.}


   \maketitle
\nolinenumbers

\section{Introduction}
\label{s:intro}

Fred Hoyle was among the first to propose the apparent angular sizes of extragalactic radio sources as standard reference objects for cosmological tests \citep{Hoyle-1959IAUS}, several years before the discovery of quasars -- at a time when the observable Universe was limited to a small fraction of the volume explored today. Practical efforts to apply Hoyle’s concept began in the 1970s with the advent of aperture synthesis in radio astronomy, which allowed imaging of radio sources with angular resolutions on the order of one arcsecond.

The idea was later formalized into the angular size–redshift ($\theta-z$) test framework by \citet{Sandage-1961ApJ}, who provided one of the earliest comprehensive treatments of such cosmological probes. As explained by the author, the $\theta-z$ test was not practical in optical astronomy, since the size of a galaxy is determined by the point at which it blends into the background, which Sandage referred to as the isophotal size. \citet{Hoyle-1959IAUS} proposed that the component separation of double-lobed radio galaxies could provide a true metric size. However, this approach proved ineffective because the intrinsic size of such sources evolves with redshift \citep[e.g.][]{Kapahi-1987IAUS}. Compact radio sources are instead motivated by their young age and location within a galaxy, suggesting their intrinsic size does not evolve with redshift \citep{KIK-1993Nature}.

A variety of cosmological tests have been developed to evaluate the expansion of the Universe. A detailed review of the most well-established methods used to probe the cosmic acceleration, Type Ia supernovae, baryon acoustic oscillations, weak gravitational lensing, and galaxy cluster abundance, as well as several alternative approaches, can be found in \citet{Weinberg:2013agg}. A quick summary of various cosmological tests is presented by \citet{lopez-corredoira-2015}, while \citet{RevModPhys.84.1151} provides a narrative account of the scientific journey throughout the 20th century that led to the discovery of the accelerating Universe. 
An important category of cosmological tests involves standard candles, such as Type Ia supernovae \citep{1998AJ....116.1009R}, which serve as powerful tools for determining the Universe's expansion history and were instrumental in the discovery of its accelerated expansion. 
The CosmoVerse White Paper \citep{VALENTINO2025101965} presents a synthesis of current constraints on key cosmological parameters, based on a comprehensive review of a broad range of probes, including the cosmic microwave background (CMB) anisotropies, large-scale structure, Type Ia supernovae, and gravitational lensing.

The relationship between the angular size of double-lobed radio sources and their redshift was first explored by \citet{MILEY1968}, who analysed 72 quasi-stellar sources at 408~MHz. Shortly thereafter, \citet{Bash1968} studied 234 sources at 2695~MHz and reported no significant relation between redshift and observed mean visibilities used to define sources' angular sizes. 
Subsequent analysis by \citet{LEGG1970} revealed that sources with higher redshifts tend to exhibit smaller angular sizes, consistent with a cosmological interpretation only if evolution of the source size with redshift is considered. 
This line of investigation was continued by \citet{Miley-1971MNRAS}, who confirmed the angular diameter--redshift relation using a sample of 127 quasars. Later, \citet{Kellermann-1972AJ} reported a weak dependence of flux density on angular size, although the result was inconsistent with predictions from simple cosmological models. 
These findings suggest that, in addition to cosmological expansion, intrinsic properties of radio sources or evolutionary effects may play a significant role in shaping the observed angular size–redshift relation. 

Subsequently, \citet{Singal-1993MNRAS} demonstrated intrinsic differences between radio galaxies and quasars. A direct correlation between luminosity and the size of radio galaxies: more luminous radio galaxies tend to have larger physical sizes, while their sizes decrease significantly with increasing redshift, indicating notable cosmic evolution. 
On the other hand, for quasars, an inverse correlation was observed: more luminous quasars generally have smaller physical sizes, exhibiting minimal evolution with redshift and suggesting a different evolutionary behaviour compared to radio galaxies.
\citet{Daly-1994ApJ} presented a model for the propagation of the radio lobes of powerful extended radio sources, allowing the estimation of intrinsic physical variables from observed data, providing insights into the dynamics of radio source propagation. Moreover, by comparing intrinsic source sizes with measured angular sizes -- where the former are nearly independent of the deceleration parameter $q_0$ -- the allowed range of $q_0$ can be constrained. The findings suggest that a low value of $q_0$ is favoured, implying that either space curvature or a cosmological constant may play a significant role.

Later, \citet{Buchalter+1998ApJ} studied a sample of 103 double-lobed quasars and found that they have no intrinsic size evolution with redshift, suggesting that the observed size changes are primarily due to cosmological effects rather than intrinsic source evolution. They concluded that a flat, matter-dominated Universe ($\Omega_0 = 1$), a flat Universe with a cosmological constant, and an open Universe all provide comparably good fits to the data. They also examined the values of the Hubble constant $H_0$ and obtained values consistent with the accepted range at that time. However, they emphasized that the sample size needed to be significantly larger for such studies to provide more robust conclusions in the future, a need addressed by the current work.

These findings collectively suggest that arcsecond-scale radio structures of extragalactic sources, such as radio galaxies and quasars, are significantly influenced by intrinsic evolution and complex astrophysical effects as standard rods for cosmological applications. 
This conclusion paved the way for new studies in the early 1990s, focusing on the more compact, milliarcsecond-scale structures of active galactic nuclei (AGN), which can be probed with very long baseline interferometry (VLBI). Such compact features offer a promising alternative for cosmological tests, as discussed, for example, by \cite{Daly+Djorgovski-2007NuPhS}.

\citet{KIK-1993sara,KIK-1993Nature} analysed the dependence on the redshift of the apparent angular size of compact structures (i.e., core--jet structures imaged with VLBI) of 78 quasars and came to the conclusion that the test does not contradict the flat Universe model. A subsequent analysis by \citet{1995MNRAS.277..753D} reinforced these findings by testing the angular size–redshift relation with compact radio sources. They emphasized the need for significantly larger samples -- on the order of at least 600 sources with comparable redshift coverage and measurement precision -- in order to robustly discriminate between competing cosmological models.
Further support came from \cite{Wilkinson+1998ASSL}, who conducted a dedicated $\theta-z$ analysis using a VLBI sample of compact sources, confirming the cosmological utility of milliarcsecond-scale structures in contrast to larger, more evolution-prone radio lobes.

\citet{LIG-1993sara,LIG-1994ApJ} investigated the $\theta-z$ dependence for 337 quasars and other types of extragalactic sources observed in the non-imaging VLBI survey by \citet{Preston+1985AJ} using two different approaches. \citet{Jackson+Dodgson-1997MNRAS} revisited the $\theta-z$ relation using a refined sub-sample of 256 ultracompact radio sources from the sample of \cite{LIG-1994ApJ}, finding that the data strongly disfavour an Einstein--de Sitter Universe and instead support a low-density, flat model with a significant cosmological constant. In a related analysis, \citet{Jackson+Dodgson-1996MNRAS} showed that such sources behave as standard rods and that models with $\Omega_{\mathrm{m}} \ll 1$ and $\Lambda \neq 0$ yield significantly better fits to the observed $\theta-z$ relation.

\citet{LIG-KIK-SF-1999} collected $5$-GHz VLBI contour maps of $330$ compact AGN in a wide range of redshifts ($0.011 \le z \le 4.72$) from the literature. They found that the data were generally consistent with standard Friedmann--Robertson--Walker cosmologies with $q_0 \lesssim 0.5$ and $\Lambda=0$, without the need to consider evolution or selection effects. However, the sample was not sufficiently large and unbiased for considering more general cosmological models, although \citet{LIG-KIK-SF-1999} published their data, triggering a series of follow-up investigations.
The above-mentioned dataset of 330 compact AGN has been extensively employed to constrain the $\Lambda$CDM paradigm and a variety of extensions involving cosmic curvature, quintessence dark energy models, interacting and dynamical dark energy, and deviations from general relativity such as $f(R)$ gravity and brane-world frameworks \citep{Lima+Alcaniz-2000GReGr, Lima+Alcaniz-2000AA, Lima+Alcaniz-2002ApJ, Sahni+Starobinsky-2000IJMPD, Chen+Ratra-2003ApJ, Qi+2017EPJC, Zheng+2017JCAP, Li+2017EPJC, Ma+2017EPJC, Xu+2018JCAP, Qi+2019MNRAS, Lian+2021MNRAS}

In addition, combined analyses with Type Ia supernovae, \ion{H}{ii} galaxy standard candles, and the ultraviolet--X-ray quasar distance relation have been conducted to study variable–$\Lambda$ scenarios, test the cosmic distance duality and Etherington relation, and expansion-rate measurements from cosmic chronometers (see e.g. \citet{Vishwakarma-2001CQGra, Li+Lin-2018MNRAS, Wan+2019PDU, He+2022ChJPh, Yang+2024arXiv}).

\cite{Gurvits-2003ASPC} provided a review of the angular size–redshift ($\theta-z$) test and discussed the observational and theoretical issues known at that time. Collectively, these studies confirm and enhance the cosmological utility of this dataset through improved selection and model testing.
\citet{Jackson-2004JCAP} reanalysed the \citet{LIG-KIK-SF-1999} data with improved methodology, strengthening the case for ultracompact radio sources as standard rods, and a follow-up analysis \citep{Jackson-2008MNRAS} confirmed their suitability for angular-diameter cosmology.
\citet{Cao+2018EPJC}, employing multi-frequency VLBI observations, further refined the sample selection and calibration of intrinsic source sizes, improving the reliability of cosmological constraints from high-redshift quasars.
\citet{Cao:2017ivt} analyzed compact radio quasars observed with milliarcsecond (mas) resolution as standard rulers for cosmological studies. Using a carefully selected sample of 120 compact radio sources spanning $0.46 < z < 2.76$, with negligible dependence on redshift and intrinsic luminosity ($|n| \simeq 10^{-3}, |\beta| \simeq 10^{-4}$), this study validates these objects as reliable standard rulers for probing the expansion history of the Universe.

In this paper, we revisit the classical angular size--redshift test using mas radio structures in AGN, based on new VLBI observations of more than $4\,000$ sources -- an order of magnitude larger than any sample used in previous studies. For the first time in over two decades, we present a significantly expanded X-band VLBI dataset, more than ten times larger than that used by \citet{LIG-KIK-SF-1999}, enabling a comprehensive re-examination of radio quasars as cosmological probes. We intentionally apply no restrictive selection cuts, and we make the full dataset publicly available, ensuring wide accessibility for future studies aiming to extract a standard ruler from compact radio sources. Our analysis demonstrates that simply increasing the sample size is not, on its own, sufficient to establish these objects as standard rulers due to the presence of several parameter degeneracies, which we quantify and discuss. Although this work provides the largest dataset of compact radio sources to date, deriving robust cosmological constraints will require additional assumptions and further methodological advances, which are beyond the scope of the present paper.

In \autoref{s:data}, we present the sample composition and explain the angular size definition, while in \autoref{s:methodology}, we describe how the observational data are compared to theoretical models. In \autoref{s:likelihood}, we introduce the statistical approach for estimating the parameters of the phenomenological model describing the source size, and in \autoref{s:results}, the results of our analysis are presented. \autoref{limitations} highlights the capabilities and limitations of our method for the $\theta-z$ test. Finally, in \autoref{s:cncl-otl}, we provide conclusions and an outlook for future studies.

\section{The observational data }
\label{s:data}

In this section, we provide a comprehensive explanation of the method used to form the data sample, beginning with a detailed description of the dataset and the criteria applied during the selection process.

\subsection{AGN VLBI size measurements}
\label{s:size_measurements}

Our analysis is based on VLBI observations of jetted AGN at X-band (frequencies around $8$~GHz, $7.6~\mathrm{GHz} < \nu_\mathrm{obs} < 8.7~\mathrm{GHz}$) compiled in the Astrogeo database\footnote{\url{ https://doi.org/10.25966/kyy8-yp57}} \citep{Astrogeo}. We rely on the measured interferometric visibilities and do not analyse the corresponding restored images. The Astrogeo database mainly collects data from geodetic VLBI observations \citep{Petrov2009}, the VLBA\footnote{Very Long Baseline Array of the National Radio Astronomy Observatory, Socorro, NM, USA} Calibrator Surveys (VCS; \citealt{Beasley2002,Petrov2005,Petrov2006,Kovalev2007, Petrov2008}), the Research \& Development -- VLBA project (RDV, \citealt{PushkarevKovalev2012,Piner2012}, the MOJAVE VLBA program (\citealt{Lister2018} and references therein), and other VLBI networks including the EVN\footnote{European VLBI Network}, LBA\footnote{Australian Long Baseline Array}, and GMVA\footnote{Global Millimeter VLBI Array} \citep{Helmboldt2007, Lee2008, Petrov2011_1, Petrov2011_2, Petrov2011_3, Petrov2012, Petrov2013, Schinzel2015, Shu2017, Jorstad2017, Petrov2019, Nair2019, Petrov2021, Popkov2021}. For our analysis presented here, we selected the X-band, because this is where observations for the largest number of AGN are available. The full dataset contains 17\,438 sources observed from 1994 to 2022, comprising more than $100\,000$ individual observations. Other frequency bands, or combinations of them, could also be used for similar studies in the future. 
The spectroscopic redshifts of the $4\,825$ AGN considered in this study at X-band were collected from the NASA/IPAC Extragalactic Database (NED)\footnote{\url{https://ned.ipac.caltech.edu/}}.

We fit the source structure at each individual observation with a simple model consisting of two circular Gaussian components. Typically, the brighter component is interpreted as the ``core'' emission, and the other one as the more extended jet emission. In this work, we use the distance between the two components as the characteristic angular size $\theta$ of an AGN; this does not require deciding which component is the ``core''. 
We utilize a fully automatic Bayesian model fitting approach, following \cite{2022ApJS..260....4P,2022MNRAS.515.1736K}, which provides uncertainty estimates of all component properties. Although these formal uncertainties are fundamentally underestimated (calibration and self-calibration effects are not accounted for), we find them directly useful for our subsequent analysis. To obtain more realistic error estimates, we increased the error bars to $10\%$ of the measured source size whenever the formal errors were smaller ($\sigma_\theta < 0.1\,\theta$).
A two-component model is clearly a simplification of the true source structure, but it captures a dominant spatial scale. We also note that the core--jet sample is contaminated by some objects that have no Doppler-boosted core--jet but compact symmetric structures \citep[e.g.][]{Wilkinson1994,AnBaan2012,Kiehlmann2024}. We did not attempt to filter them out here, since confirming the compact symmetric object (CSO) nature of individual candidate sources is a difficult task that requires observations at multiple frequencies, epochs, and even wavebands \citep[e.g.][]{Peck2000,Sokolovsky2011,Tremblay2016,Kiehlmann2024,An2025}. The fraction of CSOs is expected to be below $10\%$ \citep{Kiehlmann2024}.

\subsection{The data sample}
\label{ss:sampling}
To compose a $\theta-z$ sample for cosmological applications, we initially used all calibrated archival X-band VLBI data on AGN available in the Astrogeo database to date. This dataset comprises results measured independently by various observers using different calibration techniques. In case of multiple data calibrations available for a given VLBI measurement, we took the median value of all fitted parameters obtained from data calibrated by different observers (source size, flux density, and their uncertainties), to avoid repeated measurements for the same source entering the sample.  
When observations of a given source were available at multiple epochs, the angular sizes of AGN were measured separately for each observing epoch. Sources with angular sizes below the resolution limit of the interferometer \citep{resolution-limit} were excluded from further analysis. Subsequently, for each source, the median angular size across all epochs was calculated. After applying all filtering criteria, the final sample consists of $4\,825$ sources. 

Before the subsequent analysis, we applied the following additional filters to the dataset. First, we removed 157 sources where the formal error exceeded the measured size ($\sigma_\theta > \theta$), as these data points are insignificant and thus unreliable. Then we discarded $454$ sources from the sparsely-sampled redshift range $z<0.1$. The physical motivation is that sources at the lowest redshifts have $4-5$ orders of magnitude lower luminosities than those at high redshift in a flux-density-limited sample, making their treatment together as ``standard'' objects difficult. A similar approach was used earlier by e.g. \citet{LIG-1994ApJ,Jackson+Dodgson-1997MNRAS,Jackson-2004JCAP} with even larger cut at redshift 0.5.

\begin{figure}
    \centering
    \includegraphics[width=0.9\linewidth]{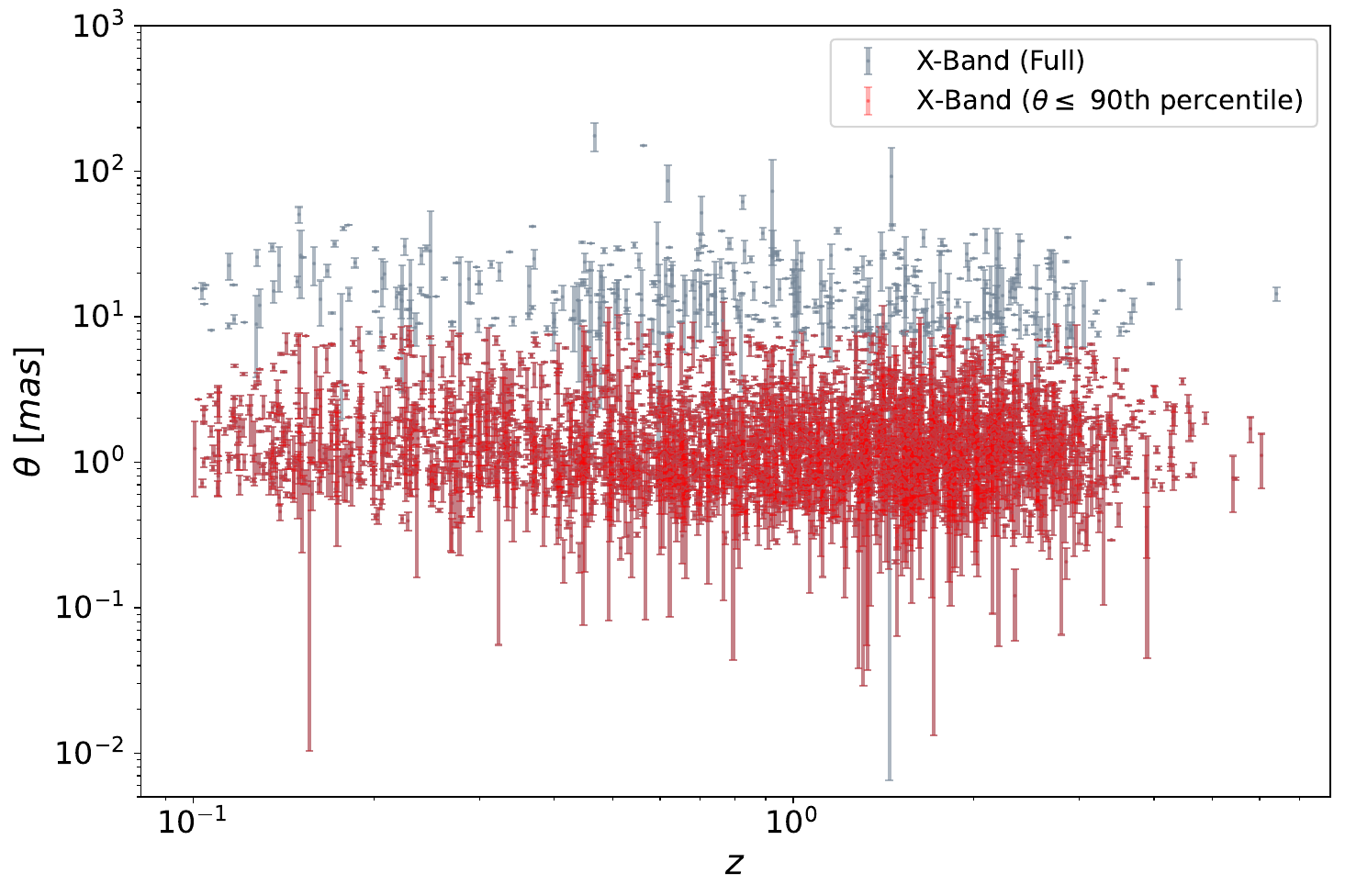}
    \caption{ Measured angular sizes (i.e. angular separation between core and jet Gaussian components fitted to the VLBI visibility data) of the AGN at X-band ($\nu_\mathrm{obs} \approx [7.6-8.7]$~GHz) in milliarcseconds versus redshift. Each dot represents one source with a median size calculated over all observing epochs together with its error bar, with the filters described in \autoref{ss:sampling} applied. The red points highlight the compact subset, defined as sources with angular sizes at or below the 90th percentile of the overall X-band size distribution.}
    \label{fig:source_sizes}
\end{figure}

The full dataset (i.e., without the low-redshift cut) underlying this study is available as an online supplementary material. The first 10 rows of the complete catalogue are shown in \autoref{tab:observation-data} as examples, while \autoref{fig:source_sizes} shows the measured angular sizes after applying the filtering criteria described above, together with a highlighted subset of compact sources defined as those with angular sizes at or below the 90th percentile of the X-band size distribution (see \autoref{s:results}).

\begin{table}[ht]
\centering
\caption{The source catalogue for the angular size--redshift test. \label{tab:observation-data}}
\small
\begin{tabular}{c|c|c|c|c|c}
\hline
Source & $z$ & $S$~[mJy] & $\sigma_S$~[mJy] & $\theta$~[mas] & $\sigma_\theta$~[mas] \\
\hline \hline
J0000$+$0307 & 2.35 & 76.7 & 0.8 & 2.04 & 0.07 \\
J0000$-$3221 & 1.28 & 251.2 & 1.4 & 6.16 & 1.20 \\
J0001$+$1456 & 0.40 & 53.1 & 1.3 & 1.69 & 3.69 \\
J0001$+$1914 & 3.10 & 241.4 & 0.7 & 0.63 & 0.05 \\
J0001$+$2358 & 0.08 & 64.7 & 1.0 & 4.04 & 3.58 \\
J0001$-$1551 & 2.04 & 167.4 & 0.9 & 1.84 & 0.04 \\
J0002$+$3032 & 2.39 & 73.7 & 2.4 & 9.60 & 2.55 \\
J0003$+$0717 & 0.12 & 18.4 & 0.5 & 1.01 & 0.13 \\
J0003$+$1210 & 0.76 & 49.3 & 1.3 & 9.44 & 6.28 \\
J0003$+$2129 & 0.45 & 151.3 & 0.8 & 1.79 & 1.01 \\
\hline
\end{tabular}
\\
Notes. The first 10 rows are shown here for guidance regarding their form and content; the table is published in its entirety at the CDS. Column description: (1) Source identifier (J2000 coordinates); (2) Spectroscopic redshift; (3) Flux density $\mathrm{[mJy]}$; (4) Flux density formal uncertainty $\mathrm{[mJy]}$ (5) Angular size $\mathrm{[mas]}$; (6) Angular size formal uncertainty $\mathrm{[mas]}$. \\ 
\end{table}

\section{Methodology}
\label{s:methodology}

In this section, we present a detailed comparison of the theoretical predictions of the flat $\Lambda$CDM model for the angular size--redshift relation of compact radio sources with the observational data described earlier. This analysis serves as the foundation for our cosmological model fitting, enabling a rigorous evaluation of the models' compatibility with the observed phenomena.

The observed angular size, $\theta (z)$, of a source at redshift $z$ can be related to its physical (linear) size, $l_\mathrm{m}$, through the angular diameter distance, $D_\mathrm{A}(z)$, via the standard expression:
\begin{equation}
    \theta (z) = \dfrac{l_\mathrm{m}}{D_\mathrm{A}(z)}.
    \label{Eq:theta-z}
\end{equation}
In the context of a flat $\Lambda$CDM Universe, we express distances in terms of the transverse comoving distance, $D_\mathrm{M}(z)$, which is equivalent to the comoving and proper distances in such the cosmological model. The relation between angular diameter distance and transverse comoving distance is given by \citep[e.g.][]{Hogg:1999ad}
\begin{equation}
D_\mathrm{A}(z) = \dfrac{D_\mathrm{M}(z)}{1+z} = \dfrac{1}{1+z} \dfrac{c}{H_0} \int^{z}_{0} \dfrac{dz'}{\sqrt{\Omega_{\mathrm{m}} (1+z')^3 + \Omega_\Lambda}}, 
\end{equation}
ensuring consistency with the cosmological background assumed. Here, $H_0$ is the Hubble constant, $c$ the speed of light, and $\Omega_{\mathrm{m}}$ the dimensionless matter density parameter. The dimensionless dark energy density parameter is given by $\Omega_\Lambda = 1 - \Omega_{\mathrm{m}}$.

Based on the phenomenological model proposed in \citet{LIG-1994ApJ, LIG-KIK-SF-1999},
\begin{equation}
l_\mathrm{m} = l \left( \frac{L}{L_0} \right)^{\beta} (1 + z)^{n}.
\label{Eq:l_m}
\end{equation}
Here, $l$ is the linear size scaling factor, $L$ the luminosity\footnote{While traditionally referred to as luminosity, this quantity is in fact the power per unit frequency interval, i.e. power density or monochromatic luminosity.}, $L_0$ the normalizing luminosity set to $10^{28}$~W\,Hz$^{-1}$, $\beta$ parametrizes the physical properties of the compact radio-emitting regions, and $n$ describes the cosmological evolution of the linear size with redshift. 
The luminosity is computed from the observed flux density, $S$, as follows:

\begin{equation}
    L = \dfrac{4 \pi \, S \, D_\mathrm{L}^2}{(1+z)^{1 + \alpha}},
\label{Eq:luminosity}
\end{equation}
where luminosity distance, $D_\mathrm{L}$, is related to the angular diameter distance, $D_\mathrm{A}$, through the distance duality relation,
\begin{equation}
D_\mathrm{L}(z) = (1+z)^2 D_\mathrm{A}(z).
\end{equation}
The factor $(1+z)^{1 + \alpha}$ results from the k-correction, which accounts for the redshifting of the observed frequency and the spectral dependence of the source emission. In the present analysis, we assume $\alpha = 0$, which is a standard assumption for compact VLBI cores that typically exhibit flat radio spectra.

For completeness, we present in \autoref{s:appendix} an alternative formulation of the $\theta-z$ relation directly in terms of the observed flux density. While the luminosity-based parametrization mentioned above follows the standard approach in the literature, the flux-density-based approach avoids the intermediate conversion via the luminosity distance and provides a more direct method of analysis.

\section{Likelihood analysis and validation tests}
\label{s:likelihood}

To constrain the parameters of the phenomenological model described in \autoref{Eq:l_m}, we adopt a Markov chain Monte Carlo (MCMC) approach \citep{mcmc}. For the cosmological background, we fix the Hubble constant to $H_0 = 70~\mathrm{km}\,\mathrm{s}^{-1}\,\mathrm{Mpc}^{-1}$ and the matter density parameter to $\Omega_{\mathrm{m}} = 0.3$, consistent with the standard flat $\Lambda$CDM cosmology. Under these assumptions, we leave the model parameters $l$, $\beta$, and $n$ free to vary. We impose uniform priors over the parameter space as $l \in [0.01, 50]$~pc, $\beta \in [-1, 1]$, and $n \in [-5, 5]$. The goal is to determine the best-fitting set of parameters that minimises the discrepancy between the model-predicted angular sizes and the observed values. 

The quality of the fit is evaluated via the $\chi^2$ statistic:
\begin{equation}
\chi^2 = \sum_{i=1}^{N} \left( \frac{\theta_{\mathrm{obs},i} - \theta_{\mathrm{th},i}}{\sigma_i} \right)^2,
\end{equation}
where $\theta_{\mathrm{obs},i}$ and $\theta_{\mathrm{th}, i}$ are the observed and theoretical angular sizes for the $i$-th source, respectively, and $\sigma_i$ denotes the associated measurement uncertainty. The summation extends over all sources in the sample. The optimal model corresponds to the parameter combination that minimises $\chi^2$. 

Given the multidimensional nature of the parameter space, we employ the MCMC method to efficiently sample the posterior distributions. This technique performs a stochastic exploration of the parameter space, starting from an initial guess and generating a chain of states that reflects the likelihood-weighted probability distribution, subject to the imposed priors.

We perform the MCMC analysis using the \texttt{Cobaya} framework \citep{2019ascl.soft10019T,Torrado:2020dgo}, which provides a flexible and efficient environment for Bayesian parameter inference. \texttt{Cobaya} interfaces with advanced samplers and allows for customized likelihood functions, making it well-suited for our model. The sampling explores the posterior distributions of the parameters $l$, $\beta$, and $n$, constrained by the observational data and the previously described priors.

\subsection{Mock catalogue construction for validating parameter recovery}
\label{ss:mock-cat}

In order to test the reliability of our MCMC pipeline, we first generated a series of mock catalogues. These were constructed by assuming a flat $\Lambda$CDM cosmology with $\Omega_{\mathrm{m}} = 0.3$ and $H_{0} = 70~\mathrm{km}\,\mathrm{s}^{-1}\,\mathrm{Mpc}^{-1}$, and a fiducial standard rod of comoving length $l_\mathrm{m} = 5\, {\rm pc}$, while $\beta = n = 0$. The corresponding angular size as a function of redshift was obtained by computing the angular diameter distance, which was implemented numerically with a dense interpolation grid to ensure computational efficiency. 

The mock measurements of the angular size were then perturbed with Gaussian noise at different relative levels, namely $10\%$, $20\%$, and $50\%$. For each case, the statistical uncertainties were taken to be proportional to the true signal, i.e. $\sigma_{\theta} = \epsilon\,\theta$ with $\epsilon = 0.1,\,0.2,\,0.5$. This allowed us to mimic semi-realistic observational conditions and assess the robustness of the parameter recovery as a function of measurement precision. 
In order to keep the analysis simple and avoid the additional complexity introduced by luminosity correlations in the real data, we set $\beta = 0$ for the rest of this test. We also tested the robustness of our pipeline by incorporating luminosity information; however, since the primary goal of this section is to validate the angular-size and cosmology inference, the details of the luminosity-based tests are not discussed here.

\subsection{Randomization test for the physical $\theta-z$ dependence}
\label{ss:random}

To further assess whether the constraints on the model parameters truly stem from the physical connection between the observable quantities, we performed an additional test using a randomized version of the dataset. In this test, we retained the observed luminosities, angular sizes, and their associated uncertainties, but randomly shuffled the redshift values among the sources. This process effectively destroys any physical correlation between redshift and the other observables, while preserving the individual distributions of each quantity. We then applied the same MCMC fitting procedure to this randomized catalogue. The outcome of this test, discussed in \autoref{ss:random-conclusion}, highlights the significance of the angular size–redshift dependence in the real data.

\section{Results of parameter estimations} 
\label{s:results}

\begin{figure}[ht]
    \centering
    \includegraphics[width=0.9\linewidth]
    {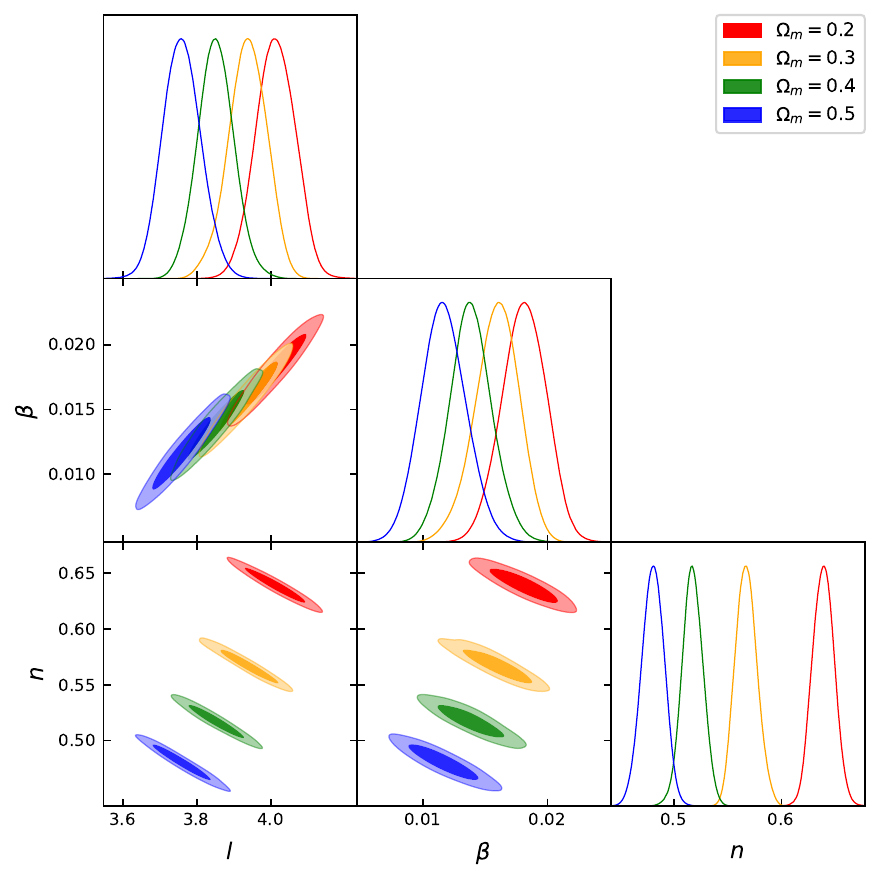}
    \caption{Corner plot showing the marginalized posterior distributions of the model parameters $l$, $\beta$, and $n$ derived from MCMC runs using the \texttt{Cobaya} package for a fixed flat $\Lambda$CDM cosmology with $H_0 = 70~\mathrm{km}\,\mathrm{s}^{-1}\,\mathrm{Mpc}^{-1}$ and four values of the matter density parameter: $\Omega_{\mathrm{m}} = 0.2$ (red), $0.3$ (orange), $0.4$ (green), and $0.5$ (blue). The contours indicate the 68\% and 95\% confidence levels. A clear trend is observed across all three parameters as $\Omega_{\mathrm{m}}$ increases: $l$, $\beta$ and $n$ systematically decrease, with the posteriors showing negligible overlap, reflecting high sensitivity to the assumed cosmological background.}
    \label{fig:corner_real}
\end{figure}

After applying the MCMC analysis to the real dataset (\autoref{tab:observation-data}), the resulting posterior distributions of the parameters $l$, $\beta$, and $n$ for each cosmological case ($\Omega_{\mathrm{m}} = 0.2$, $0.3$, $0.4$, and $0.5$) are presented in \autoref{fig:corner_real}. The corresponding best-fitting values with their $1\sigma$ uncertainties are summarized in \autoref{tab:real_results}.

\begin{table}
    \centering
    \caption{Mean values and $1\sigma$ uncertainties for the parameters \( l \), \( \beta \), and \( n \) at X-band for $\Omega_{\mathrm{m}} = 0.2$, $0.3$, $0.4$, and $0.5$.}
    \label{tab:real_results}
    \begin{tabular}{c|c}
        \hline
        Parameter & Value \\ 
        \hline \hline
        \multicolumn{2}{c}{\textbf{$\Omega_{\mathrm{m}} = 0.2$}} \\ 
        \hline
        \( l ~ \mathrm{[pc]}\)      & \( 4.013 \pm 0.053 \) \\
        \( \beta \)  & \( 0.018 \pm 0.002 \) \\
        \( n \)      & \( 0.639 \pm 0.011 \) \\  
        \hline
        \multicolumn{2}{c}{\textbf{$\Omega_{\mathrm{m}} = 0.3$}} \\  
        \hline
        \( l ~ \mathrm{[pc]}\)      & \( 3.937 \pm 0.052 \) \\
        \( \beta \)  & \( 0.016 \pm 0.018 \) \\
        \( n \)      & \( 0.567 \pm 0.010 \) \\  
        \hline
        \multicolumn{2}{c}{\textbf{$\Omega_{\mathrm{m}} = 0.4$}} \\  
        \hline
        \( l ~ \mathrm{[pc]}\)      & \( 3.852 \pm 0.049 \) \\
        \( \beta \)  & \( 0.014 \pm 0.002 \) \\
        \( n \)      & \( 0.517 \pm 0.010 \) \\  
        \hline
        \multicolumn{2}{c}{\textbf{$\Omega_{\mathrm{m}} = 0.5$}} \\  
        \hline
        \( l ~ \mathrm{[pc]}\)      & \( 3.759 \pm 0.051 \) \\
        \( \beta \)  & \( 0.012 \pm 0.002 \) \\
        \( n \)      & \( 0.480 \pm 0.010 \) 
         \\  
        \hline
    \end{tabular}
\end{table}

We find that the model parameters are consistently well-constrained across different cosmologies. In particular, the parameter $n$, which characterises the redshift dependence of the intrinsic size, deviates significantly from zero, with typical values around $ n \simeq 0.48 - 0.64$, depending on $\Omega_{\mathrm{m}}$. One possible explanation is the physical evolution in the intrinsic sizes of the sources with redshift. However, we use data obtained at a fixed observing frequency ($\nu_\mathrm{obs} \approx 8$~GHz) which corresponds to different rest-frame frequencies, $\nu_\mathrm{rest}=(1+z) \nu_\mathrm{obs}$, for each individual source. Therefore, if there is any dependence of the source linear size on frequency \citep[see e.g.][for compact mas-scale radio sources]{Blandford+Rees-1978PhScr} then it should also contribute to the value of parameter $n$ in our phenomenological formula (\autoref{Eq:l_m}).   
Similar concerns were raised in earlier works (e.g. \citealt{KIK-1993Nature}), which highlighted the potential impact of evolutionary and observational effects on angular size--redshift relations in compact radio sources. 

Furthermore, the best-fitting values of $l$ and $\beta$ exhibit minimal variation with respect to the assumed matter density parameter, showing changes of only a few percent across the tested range of $\Omega_{\mathrm{m}}$. This suggests that, while the dataset effectively constrains the model parameters, it has limited sensitivity to variations in $\Omega_{\mathrm{m}}$ within the explored range. 

To assess the influence of extreme angular-size outliers on the inferred cosmological parameters, we repeated the MCMC analysis after removing the upper $10\%$ of the angular-size distribution, corresponding to sources with $\theta$ values above the 90th percentile. This cut excludes only $422$ of the $4\,214$ sources (resulting in $3792$ retained). The angular-size distribution is strongly skewed, with the vast majority of sources having $\theta < 10$~mas and only a few extreme cases reaching $\theta > 40$~mas.

The posterior mean values and covariance matrices of the parameters $n$, $l$, and $\beta$ remain nearly identical after applying this cut, with all shifts well below their respective $1\sigma$ uncertainties. This demonstrates that the large-$\theta$ tail contributes negligibly to the likelihood, and the fitted cosmological trends are instead governed by the dominant population of compact sources. The test thus confirms that the derived parameters are not driven by a small number of unusual, extended AGN, and it verifies the robustness of the results.

\subsection{Parameter recovery with mock catalogues}
\label{ss:mock-conclusion}

\begin{figure}[ht]
    \centering
    \includegraphics[width=0.9\linewidth]{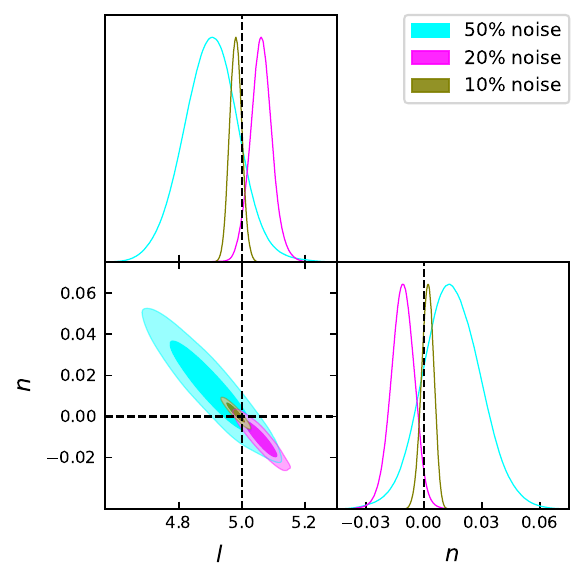}
    \caption{
    Corner plot showing the 2D posterior distributions for the parameters $l$ [pc] and $n$ derived from mock catalogues with $10\%$ (olive), $20\%$ (magenta), and $50\%$ (cyan) Gaussian noise. Dashed black lines indicate the fiducial values assumed when generating the mock catalogues. Contours show the degeneracy between $l$ and $n$, which becomes more elongated with increasing noise, reflecting larger uncertainties along the degenerate direction. Despite the increasing noise, the contours still overlap, demonstrating that the pipeline robustly recovers the fiducial values.
    }
    \label{fig:mock_10_20_50}
\end{figure}

As shown in \autoref{fig:mock_10_20_50}, our pipeline successfully recovers the fiducial values for $l$ and $n$. As expected, increasing noise broadens and enlarges the posterior contours; nevertheless, the contours still overlap, indicating that the pipeline remains robust. In this figure, the contours are shown in cyan, magenta, and olive for the $50\%$, $20\%$, and $10\%$ noise cases, respectively. The dashed black lines indicate the fiducial values assumed when generating the mock catalogues. The corresponding mean values and $1\sigma$ uncertainties are listed in \autoref{tab:mock_10_20_50s}.

The 2D posteriors exhibit the characteristic ``banana-shaped'' contours that clearly reveal the degeneracy between $l$ and $n$, as anticipated from the theoretical scaling relations. As the noise level increases, the contours become more elongated, indicating larger uncertainties along the degenerate direction, while still remaining consistent with the fiducial values.

\begin{table}[h!]
    \centering
    \caption{Mean values and $1\sigma$ uncertainties for the parameters $l$ and $n$ from mock catalogues with $10\%$, $20\%$, and $50\%$ Gaussian noise.}
    \label{tab:mock_10_20_50s}
    \begin{tabular}{c|c|c}
        \hline
        Noise & $l$ [pc] & $n$ \\
        \hline \hline
        10\% & $4.980 \pm 0.018$ & $1.75 \times 10^{-3} \pm {3.1} \times 10^{-3}$ \\
        20\% & $5.060 \pm 0.036$ & $-0.011 \pm 0.006$ \\
        50\% & $4.900 \pm 0.091$ & $0.015 \pm 0.015$ \\
        \hline
    \end{tabular}
\end{table}

These results demonstrate that the pipeline reliably recovers the fiducial input values, with uncertainties increasing in a predictable manner with noise. The overlap of the contours confirms that even at $10\%$ noise, the degeneracy between $l$ and $n$ remains constrained.

\subsection{Results of the randomization test}
\label{ss:random-conclusion}

\begin{figure}[ht]
    \centering
    \includegraphics[width=0.8\linewidth]{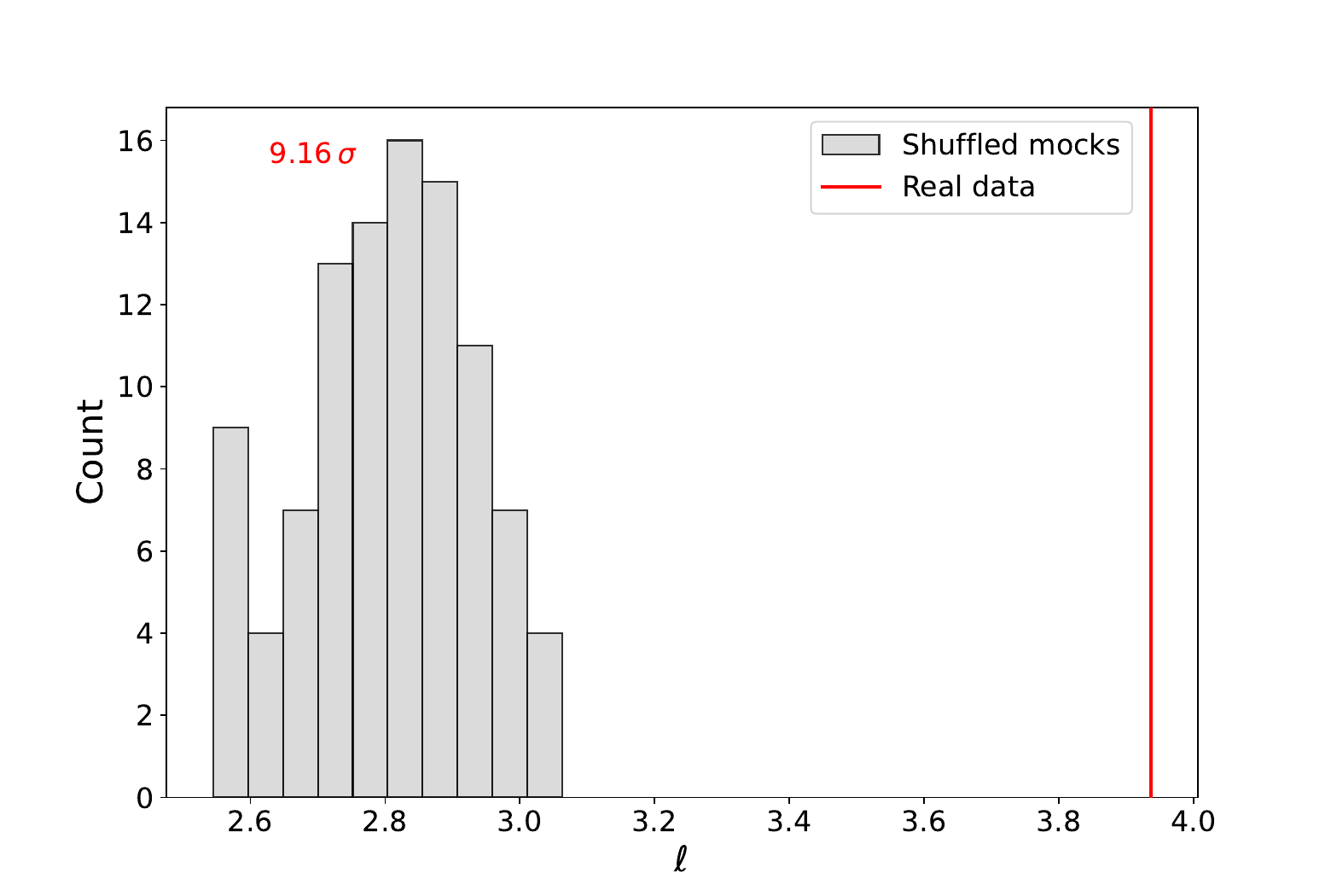}
    \includegraphics[width=0.8\linewidth]{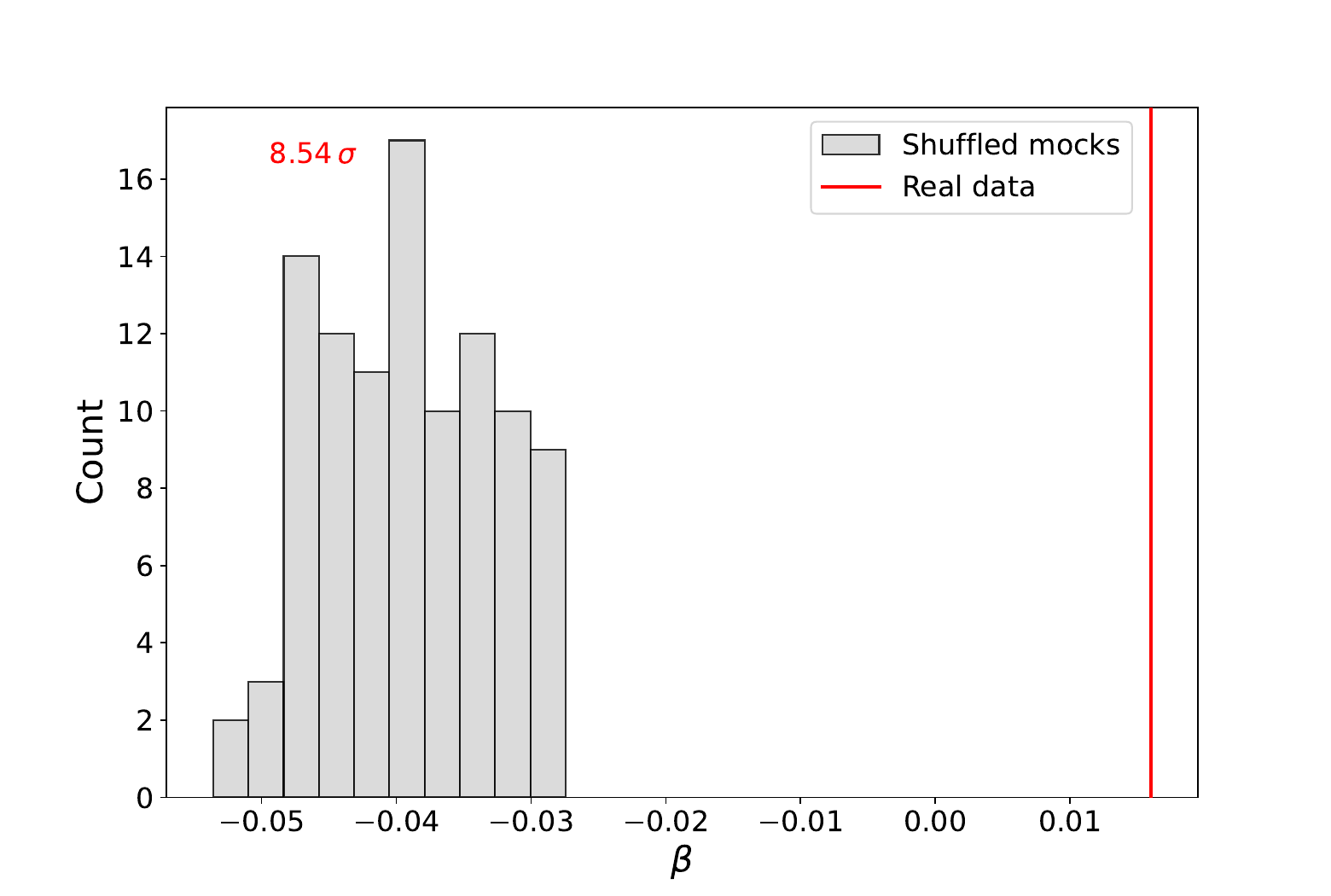}
    \includegraphics[width=0.8\linewidth]{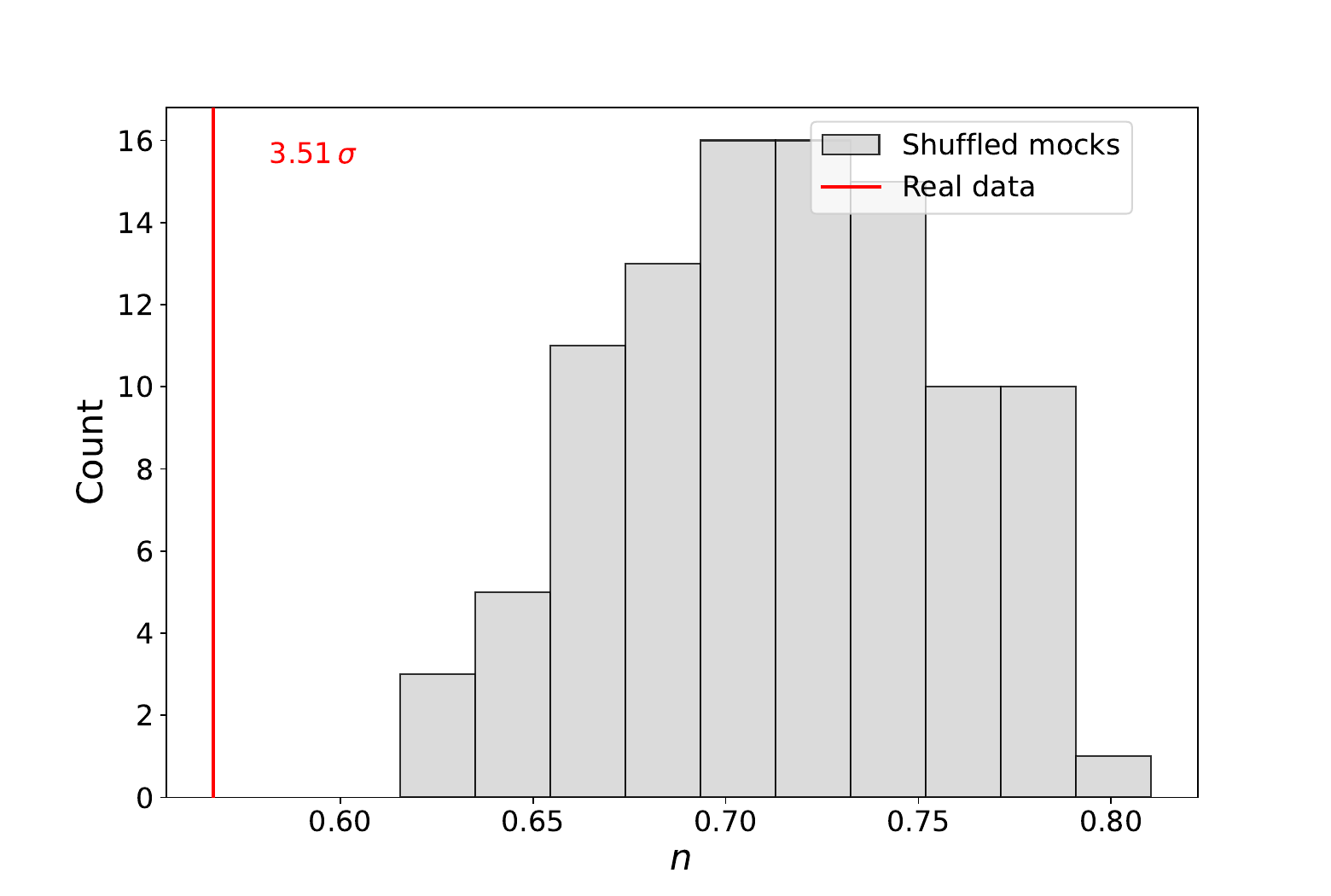}
    \caption{Distribution of the best-fit parameters $l$, $\beta$, and $n$ obtained from 100 mock catalogues with randomly shuffled redshifts. The grey histograms represent the null distributions under the assumption of no intrinsic correlation between angular size and redshift. The red vertical lines mark the values of the parameters derived from the real (unshuffled) data: $l$ (9.16 $\sigma$), $\beta$ (8.54 $\sigma$), and $n$ 
    (3.51 $\sigma$). All three show deviations from the mock distributions, indicating significant detections of the correlations.}
    \label{fig:hist_l}
\end{figure}

To assess the statistical significance of the angular size–redshift relation in our sample, we generated 100 randomized catalogues by randomly shuffling the redshifts while preserving all other observables, as described in detail in \autoref{ss:random}. For each randomized catalogue, we repeated the full MCMC analysis using the same setup $H_0 = 70~\mathrm{km}\,\mathrm{s}^{-1}\,\mathrm{Mpc}^{-1}$ and $\Omega_{\mathrm{m}} = 0.3$, thereby constructing a null distribution for each of the model parameters $l$, $\beta$, and $n$. As shown in \autoref{fig:hist_l} for $l$, $\beta$, and $n$, the real-data values lie well outside the null distributions obtained from the shuffled mocks, yielding a 9.16 $\sigma$ deviation. Similarly, we find 8.54 $\sigma$ for $\beta$ and 3.51 $\sigma$ for $n$. These significant deviations suggest that the inferred parameters are not the result of random associations but reflect a genuine underlying dependence between angular size and redshift in the observed data.

As expected, the MCMC analysis of randomized catalogues failed to constrain the model parameters, with the resulting posterior distributions being broad. This confirms that the ability to recover meaningful constraints in the real data relies on the genuine physical relationship between flux density, redshift, and angular size, rather than arising from statistical artifacts or chance alignments.

\section{Capabilities and limitations of the approach for cosmological tests}
\label{limitations}

We performed a simplified evaluation of the potentially achievable precision with the proposed approach and the conditions required to reach it. To gain a clearer understanding, it is instructive to reformulate the equations used in the analysis, from \autoref{Eq:theta-z} and \autoref{Eq:l_m}:
\begin{equation}
\theta = \frac{l}{D_\mathrm{A}} \left( \frac{4 \pi \, S \, D^2_\mathrm{L}}{(1+z)^{1+\alpha} \, L_0} \right)^{\beta} (1 + z)^{n}.
\label{Eq:combined}
\end{equation}

Now we aim to express all redshift-dependent (distance-dependent) terms in the right-hand side (RHS), and combine all measurable quantities per source in the left-hand side (LHS). This equation takes the form:
\begin{equation}
    \dfrac{\theta}{S^{\beta}} = \mathrm{const} \cdot \frac{(1+z)^n \, D_\mathrm{L}^{2 \beta}}{(1+z)^{(1+\alpha) \beta} \, D_\mathrm{A}} = \mathrm{const} \cdot (1+z)^{n + 3 \beta} \, D_\mathrm{A}^{2 \beta - 1}.
    \label{eq:lhs_rhs}
\end{equation}

The RHS is the effective dependency we aim to constrain using observational data. On the other hand, the LHS is the directly measurable quantity, responsible for the observational constraint of the equation.

\begin{figure}[ht]
    \centering
    \includegraphics[width=0.9\linewidth]{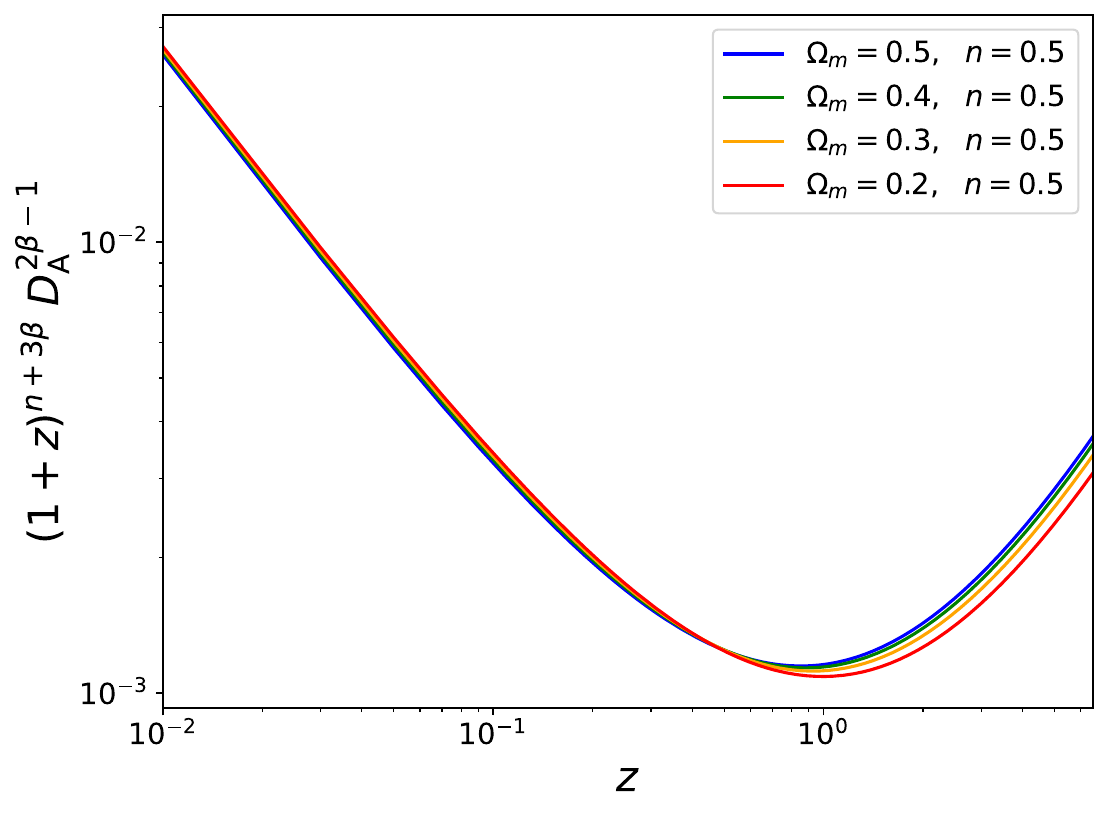}
    \includegraphics[width=0.9\linewidth]{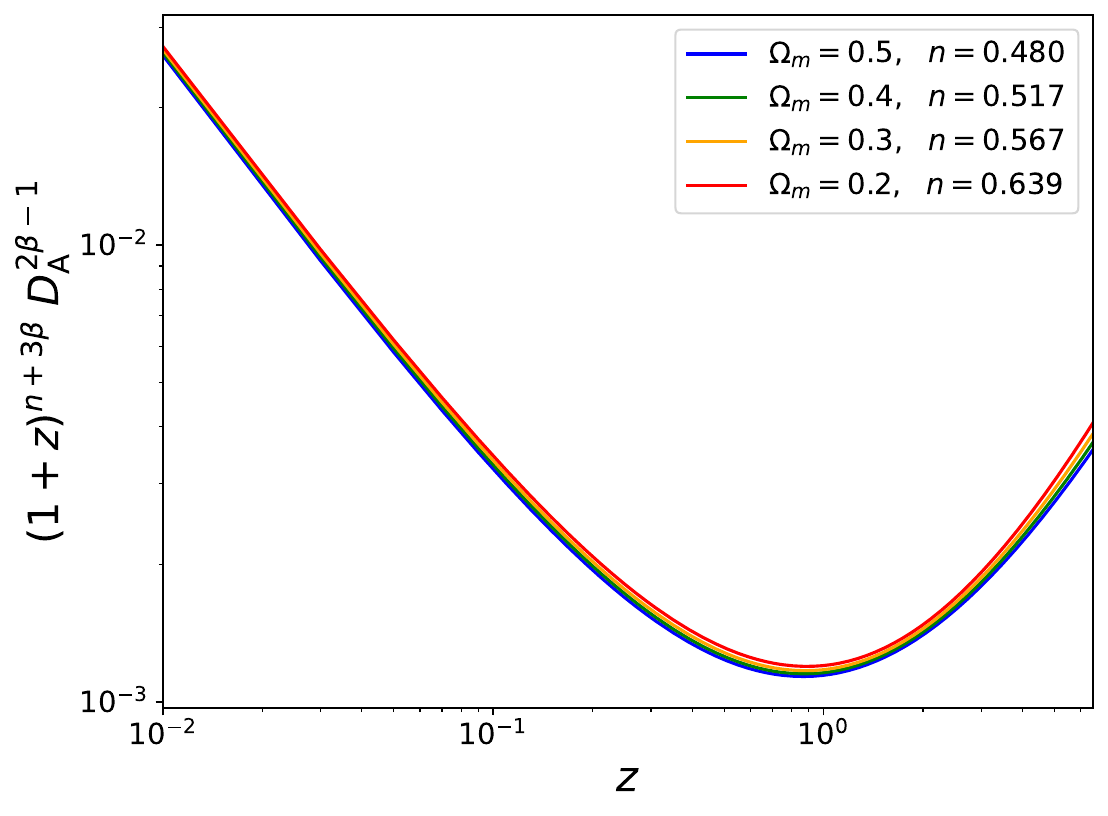}
    \caption{The redshift-dependent part of the right-hand side of \autoref{eq:lhs_rhs} plotted for different cosmological parameters; $D_A$ is measured in Mpc. Top curves for fixed $n$ and varying $\Omega_\mathrm{m} = 0.2$ (red), $0.3$ (orange), $0.4$ (green), and $0.5$ (blue) with $\beta$ fixed to the fitted values of \autoref{tab:real_results}. At higher redshifts, deviations between cosmologies become better articulated. Bottom: Both parameters, $\Omega_{\rm m}$ and $n$, change together based on mean values of \autoref{tab:real_results}, again with $\beta$ best-fit values. Adjusting $n$ effectively compensates for the differences in $\Omega_\mathrm{m}$, resulting in curves that are practically indistinguishable from each other across the full redshift range. Note that, despite a similar shape, these curves are not identical to that depicting the angular size of a hypothetical true standard metric rod as a function of redshift.}
    \label{fig:sasha_rhs}
\end{figure}

In \autoref{fig:sasha_rhs}, we illustrate the behaviour of the RHS of \autoref{eq:lhs_rhs} for different cosmological models, with $\beta$ from \autoref{tab:real_results}. In the upper panel, we fix $n$ and vary the matter density parameter $\Omega_{\mathrm{m}} = 0.2$ (red), $0.3$ (orange), $0.4$ (green), and $0.5$ (blue). At high redshift, the curves begin to diverge, with differences reflecting the underlying cosmological dependence of the angular size–redshift relation. Here, only $\Omega_{\mathrm{m}}$ varies while $n$ is held fixed.

The lower panel repeats the comparison using the same $\Omega_{\mathrm{m}}$ values, but allowing $n$ to vary according to the mean values reported in \autoref{tab:real_results}. Here, both $n$ and $\Omega_{\mathrm{m}}$ vary simultaneously. In this case, adjustments to $n$ largely compensate for the changes in $\Omega_{\mathrm{m}}$, producing curves that are almost indistinguishable across the entire redshift range. This highlights a strong degeneracy between $n$ and $\Omega_{\mathrm{m}}$, consistent with the correlations visible in the posterior distributions in \autoref{fig:corner_real}. Such degeneracy implies that precise constraints on $\Omega_{\mathrm{m}}$ require $n$ to be determined independently or fixed by external considerations.

Let us now consider an idealised scenario in which $\beta$ can be determined independently and fixed in the analysis. In such a case, the remaining uncertainties originate solely from random errors in the measured angular sizes and flux densities. 
Based on the observed scatter in the LHS versus redshift dependency, the effective per-source uncertainty is approximately a factor of two. Reaching the required sub-percent precision in the RHS thus necessitates a sample size of $(\sigma_1/\sigma_{\mathrm{tot}})^2$ sources, where $\sigma_1$ represents the uncertainty in a single measurement (per-source scatter) and $\sigma_{\mathrm{tot}}$ is the target uncertainty for the combined result. For a typical value of $\sigma_1 \approx 2$ and aiming for $\sigma_{\mathrm{tot}} \approx 0.01$, this translates to a required sample of approximately $5\,000 - 100\,000$ objects, depending on the exact precision goals. This quick calculation indicates that we are only now beginning to approach the absolute minimum number of sources necessary for this method to become cosmologically constraining. Any slight deviation from the model assumptions made in \autoref{Eq:l_m} directly affects the already subtle differences between model predictions for different cosmological parameters.

\section{Conclusions and outlook}
\label{s:cncl-otl}

We analysed a large dataset of $4\,214$ compact radio-emitting AGN with mas-scale characteristic angular sizes derived from VLBI visibility model fitting at X-band, and spectroscopic redshifts. The angular size--redshift dependence qualitatively follows the $\Lambda$CDM expectation, showing a flattening near $z \sim 1$. Using a model in which the intrinsic linear size depends on flux density and redshift (\autoref{Eq:l_m}), we performed MCMC fits for the parameters $l$, $\beta$, and $n$ under fixed flat $\Lambda$CDM cosmologies with $\Omega_{\mathrm{m}}$ in the range 0.2–0.5.

The parameter $n$ is consistently non-zero ($n \simeq 0.48 - 0.64$), caused by a combination of redshift evolution of the intrinsic source size, and its dependence on the emitted frequency which equals $(1+z)$ times the fixed observing frequency $\nu_{\mathrm{obs}} \approx 8$~GHz. 
While $l$ and $\beta$ show only minor changes with $\Omega_{\mathrm{m}}$, $n$ is strongly degenerate with $\Omega_{\mathrm{m}}$, preventing simultaneous constraints on cosmology and all astrophysical parameters. Adjustments in $n$ can mimic the effect of changing $\Omega_{\mathrm{m}}$, producing nearly identical angular size–redshift curves across the full $z$ range.

We verified the stability of our results by repeating the full analysis after excluding the $10\%$ largest angular-size sources. The resulting posterior constraints are consistent within $\ll 1\sigma$, confirming that outliers have a negligible impact on the cosmological inference. This finding indicates that our conclusions are robust against the inclusion of a small number of large, resolved sources and that the cosmological signal arises primarily from the compact AGN population that dominates the sample.

Creating mock catalogues by adding Gaussian noise to the theoretical angular size values demonstrates that the MCMC pipeline reliably recovers the assumed nuisance parameters in the absence of flux density information. These tests also highlight the intrinsic degeneracy between $l$ and $n$, which persists regardless of the realism of the data, highlighting the need for precise measurements, additional assumptions, or complementary information from independent sources to break the degeneracy between $l$ and $n$ in future applications.

Given the observed per-source scatter (factor of $\sim 2$), sub-percent precision would require $5\,000$–$100\,000$ sources under idealized conditions. Current samples are at the lower bound of this requirement.
A randomization test using 100 shuffled-redshift catalogues yields deviations up to 9.16 $\sigma$ from the real-data parameter values, confirming that the angular size–redshift correlation is not due to chance.

We conclude that the presently analysed single-frequency dataset, in its entirety, cannot be treated for joint cosmology--astrophysics inference without additional constraints. The degeneracy between $n$ and $\Omega_{\mathrm{m}}$ must be broken, for example by \textit{(1)} selecting subsamples by restricting the sources to narrow ranges of luminosity, radio spectral index, and/or rest-frame frequency, or \textit{(2)} cross-matching with multi-frequency or multi-wavelength data to model size--frequency and size--luminosity dependencies explicitly.

To facilitate follow-up studies, we publish the full X-band VLBI angular size--redshift dataset in machine-readable form. With improved source characterization and larger, high-precision VLBI samples, the $\theta-z$ test could become a competitive probe of cosmology, particularly at the high-redshift regime poorly covered by other techniques.

\section*{Data Availability}

The analysis is based on the data compiled in the Astrogeo database\footnote{\url{https://doi.org/10.25966/kyy8-yp57}} that collects VLBI observations. Details are presented in \autoref{s:size_measurements}.
\autoref{tab:observation-data} is only available in electronic form at the CDS via anonymous ftp to \url{cdsarc.u-strasbg.fr} (130.79.128.5) or via \url{http://cdsweb.u-strasbg.fr/cgi-bin/qcat?J/A+A/}.

\begin{acknowledgements} 
      We gratefully acknowledge the anonymous referee for the insightful comments and suggestions. M.G.Y. highly appreciates the valuable discussions and continuous support from Istv\'an Csabai and P\'eter Raffai. We used in our work the Astrogeo VLBI FITS image database, DOI: 10.25966/kyy8-yp57, maintained by Leonid Petrov. This research has made use of the NASA/IPAC Extragalactic Database, which is funded by the National Aeronautics and Space Administration and operated by the California Institute of Technology. M.G.Y. was supported by the EK\"OP-24 University Excellence Scholarship Program of the Ministry for Culture and Innovation from the source of the National Research, Development and Innovation Fund.
      S.F. acknowledges funding received from the Hungarian National Research, Development and Innovation Office (NKFIH), grant number OTKA K134213, and from the NKFIH excellence grant TKP2021-NKTA-64.
      The Large-Scale Structure (LSS) research group at Konkoly Observatory has been supported by a \emph{Lend\"ulet} excellence grant by the Hungarian Academy of Sciences (MTA). This project has received funding from the European Union’s Horizon Europe research and innovation programme under the Marie Sk{\l}odowska-Curie grant agreement number 101130774. Funding for this project was also available in part through NKFIH grant OTKA NN147550.
      A.V.P.\ is a postdoctoral fellow at the Black Hole Initiative, which is funded by grants from the John Templeton Foundation (grants 60477, 61479, 62286) and the Gordon and Betty Moore Foundation (grant GBMF-8273). 
      The NRAO is operated by the Associated Universities Inc. under cooperative agreement with the US National Science Foundation. 
\end{acknowledgements}

\bibliographystyle{aa}
\bibliography{theta-z}

\begin{appendix}
\section{A flux-density-based approach to computing $\theta(z)$}
\label{s:appendix}

In previous studies, the angular size–redshift relation, $\theta(z)$, has typically been formulated in terms of the intrinsic luminosity, $L$ (\autoref{Eq:l_m}), which is then connected to the observed flux density via the luminosity distance (\autoref{Eq:luminosity}). Because the luminosity distance depends on the assumed cosmological model, this approach introduces an additional layer of model dependence into the analysis. However, this intermediate step is not strictly necessary. In this Appendix, we reformulate the $\theta(z)$ relation directly in terms of the observable flux density. 
In this framework, the relation between $\theta$ and $z$ can be written as
\begin{equation}
    \theta(z) = \dfrac{r_{\rm m}}{D_{\rm A}(z)},
\end{equation}
where the linear size is
\begin{equation}
    r_{\rm m} = r \left( \dfrac{S}{S_0} \right)^{b} (1 + z)^{a}.
\label{Eq:fluxd}
\end{equation}
Here, $r$ is the linear size scaling factor, $S$ is the measured flux density, and $S_0 = 1$~Jy is the normalizing flux density. The parameter $b$ describes the physical properties of the compact radio-emitting regions, and $a$ the cosmological evolution of the linear size with redshift. (Note the structural similarity with \autoref{Eq:l_m}.) This formulation retains a fully observational foundation and eliminates reliance on assumed cosmological distances or luminosity calibrations. Note that the actual value of $r$ depends on the choice of the normalizing flux density $S_0$.

The above reformulation has two practical advantages: a) Observable-level inference: The constraints are expressed directly in terms of the observed flux density, which avoids the propagation of possible additional uncertainties associated with luminosity. b) Model consistency: Since luminosity depends on an assumed model, using the flux density ensures internal consistency and avoids circular dependencies. 
In order to assess the ability of this reformulation, we apply the same sample of $4\,214$ sources used in the main analysis (\autoref{ss:sampling}) and implement a pipeline that operates directly on the flux density (\autoref{Eq:fluxd}). All other assumptions and analysis choices are kept identical to the base pipeline described in \autoref{s:likelihood}. 

\begin{figure}[ht]
    \centering
    \includegraphics[width=0.9\linewidth]
    {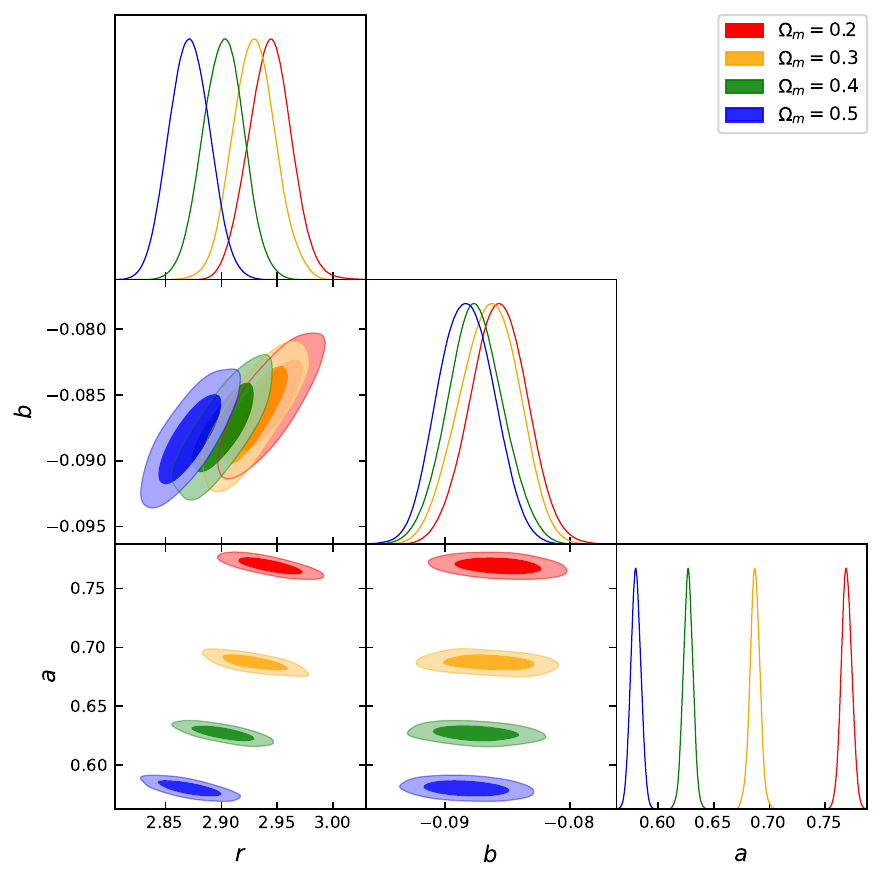}
    \caption{Corner plot showing the marginalized posterior distributions of the model parameters $r$, $b$, and $a$ derived from MCMC runs using the \texttt{Cobaya} package for a fixed flat $\Lambda$CDM cosmology with $H_0 = 70~\mathrm{km}\,\mathrm{s}^{-1}\,\mathrm{Mpc}^{-1}$ and four values of the matter density parameter: $\Omega_{\mathrm{m}} = 0.2$ (red), $0.3$ (orange), $0.4$ (green), and $0.5$ (blue). The contours indicate the 68\% and 95\% confidence levels. A clear trend is observed across all three parameters as $\Omega_{\mathrm{m}}$ increases: $r$, $b$, and $a$ systematically decrease, with the posteriors showing negligible overlap, reflecting high sensitivity to the assumed cosmological background.}
    \label{fig:corner_flux}
\end{figure}

\autoref{fig:corner_flux} presents the marginalized posterior distributions of the model parameters $r$, $b$, and $a$, for each cosmological case $\Omega_{\rm m} =$ 0.2 (red), 0.3 (orange), 0.4 (green), and 0.5 (blue).

\begin{table}
    \centering
    \caption{Mean values and $1\sigma$ uncertainties for the parameters \( r \), \( b \), and \( a \) at X-band for $\Omega_{\mathrm{m}} = 0.2$, $0.3$, $0.4$, and $0.5$.}
    \label{tab:flux_results}
    \begin{tabular}{c|c}
        \hline
        Parameter & Value \\ 
        \hline \hline
        \multicolumn{2}{c}{\textbf{$\Omega_{\mathrm{m}} = 0.2$}} \\ 
        \hline
        \( r ~ \mathrm{[pc]}\)      & \( 2.943 \pm 0.019 \) \\
        \( b \)  & \( -0.086 \pm 0.002 \) \\
        \( a \)      & \( 0.769 \pm 0.005 \) \\  
        \hline
        \multicolumn{2}{c}{\textbf{$\Omega_{\mathrm{m}} = 0.3$}} \\  
        \hline
        \( r ~ \mathrm{[pc]}\)      & \( 2.929 \pm 0.019 \) \\
        \( b \)  & \( -0.086 \pm 0.002 \) \\
        \( a \)      & \( 0.687 \pm 0.004 \) \\  
        \hline
        \multicolumn{2}{c}{\textbf{$\Omega_{\mathrm{m}} = 0.4$}} \\  
        \hline
        \( r ~ \mathrm{[pc]}\)      & \( 2.902 \pm 0.018 \) \\
        \( b \)  & \( -0.087 \pm 0.002 \) \\
        \( a \)      & \( 0.627 \pm 0.004 \) \\  
        \hline
        \multicolumn{2}{c}{\textbf{$\Omega_{\mathrm{m}} = 0.5$}} \\  
        \hline
        \( r ~ \mathrm{[pc]}\)      & \( 2.871 \pm 0.018 \) \\
        \( b \)  & \( -0.088 \pm 0.002 \) \\
        \( a \)      & \( 0.580 \pm 0.005 \) 
         \\  
        \hline
    \end{tabular}
\end{table}

We emphasise that although the formulations based on flux density and luminosity have similar expressions, the corresponding parameters $(\beta, \, n)$ and $(b,\, a)$ should not be interpreted as directly equivalent. In the luminosity-based formula, parameters encode scaling relations involving the intrinsic source luminosity, which depend on the luminosity distance. Any mapping between the two formulations requires the assumption of a cosmological model.

\end{appendix}

\end{document}